\documentclass[final]{article}

\usepackage[utf8]{inputenc}
\usepackage[T1]{fontenc}
\usepackage[english]{babel}

\usepackage{amsmath,amsthm,amssymb,amsfonts}
\usepackage{mathabx}

\newtheorem{lemma}{Lemma}
\newtheorem{theorem}{Theorem}

\newcommand{\calR}{\mathcal R}

\newcommand{\calU}{\mathcal U}
\newcommand{\calS}{\mathcal S}

\newcommand{\cond}{\;|\;}
\newcommand{\eps}{\varepsilon}

\newcommand{\E}{\mathbf{E}}
\newcommand{\xor}{\oplus}
\newcommand{\Prp}[1]{\Pr\!\left[{#1} \right]}
\newcommand{\Ep}[1]{\E\!\left[{#1} \right]}
\newcommand{\SigEps}{\Sigma^{1-\eps}}
\newcommand{\sm}{\,\setminus\,}

\newcommand\drop[1]{}
\newcommand\req[1]{(\ref{#1})}

\usepackage[pdftex]{graphicx}
\usepackage[usenames,dvipsnames,table]{xcolor}
\definecolor{shade}{RGB}{235,235,235}
\usepackage{hyperref}

\newcommand\numberthis{\addtocounter{equation}{1}\tag{\theequation}}

\newcommand*\samethanks[1][\value{footnote}]{\footnotemark[#1]}

\title{Approximately Minwise Independence with Twisted Tabulation}
\author{Søren Dahlgaard\thanks{Research partly supported by Thorup's Advanced
        Grant from the Danish Council for Independent Research under the Sapere
        Aude research carrier programme.}
        \qquad Mikkel Thorup\samethanks \vspace{8pt}\\
University of Copenhagen,\\
\tt{\{soerend,mthorup\}@di.ku.dk}}

\begin{document}

\maketitle

\begin{abstract}
A random hash function $h$ is $\eps$-minwise if for any set $S$, $|S|=n$,
and element $x\in S$, $\Pr[h(x)=\min h(S)]=(1\pm\eps)/n$. Minwise hash
functions with low bias $\eps$ have widespread applications within similarity estimation.

Hashing from a universe $[u]$, the twisted tabulation hashing of
P\v{a}tra\c{s}cu and Thorup [SODA'13] makes $c=O(1)$ lookups in tables of size
$u^{1/c}$. Twisted tabulation was invented to
get good concentration for hashing based sampling. Here we show that
twisted tabulation yields $\tilde O(1/u^{1/c})$-minwise hashing.

In the classic independence paradigm of Wegman and Carter [FOCS'79]
$\tilde O(1/u^{1/c})$-minwise hashing requires $\Omega(\log u)$-independence
[Indyk SODA'99]. P\v{a}tra\c{s}cu and Thorup [STOC'11] had shown that
simple tabulation, using same space and lookups yields
$\tilde O(1/n^{1/c})$-minwise independence, which is good for large sets, but useless for small sets. Our analysis uses some of the same methods,
but is much cleaner bypassing a complicated induction argument.

\end{abstract}

\section{Introduction}
The concept of minwise hashing (or the ``MinHash algorithm''
according to
\footnote{See \url{http://en.wikipedia.org/wiki/MinHash}} ) is a basic
algorithmic tool suggested by Broder et al.~\cite{broder97onthe,broder98minwise} for
problems related to set similarity and containment. After the initial
application of this algorithm in the early AltaVista search engine to
detecting and clustering similar documents, the scheme has reappeared
in numerous other applications\footnotemark[1] and is now a standard
tool in data mining where it is used for estimating similarity
\cite{broder98minwise,broder97onthe,broder97minwise},
rarity \cite{DatarM02estimatingrarity},
document duplicate detection
\cite{Broder00,manku07duplicates,YC06,henzinger06duplicates},
large-scale learning \cite{li11minhash},
etc.~\cite{BHP09,BPR09,CDFGIMUY01,SWA03}.

The basic motivation of minwise independence is to use hashing to select an element
from a set $S$. With a hash function $h$, we simply pick the element
$x\in S$ with the minimum hash value. If the hash function is fully
random and no two keys get the same hash, then $x$ is uniformly
distributed in $S$.

A nice aspect of minwise selection is that $\min h(A\cup
B)=\min\{\min h(A), \min h(B)\}$. This makes it easy, e.g., to select
a random leader in many distributed settings. It also
implies that that $\min h(A\cup B)\in
h(A\cap B)\iff \min h(A)=\min h(B)$.  Therefore, if $h$ is fully random and
collision free,
\[
    \Pr_{h}[\min h(A) = \min h(B)] = \frac{|A\cap B|}{|A\cup
    B|}\enspace .
\]
Thus, if we, for two
sets $A$ and $B$, have stored $\min h(A)$ and $\min h(B)$, then
we can use $[\min h(A) = \min h(B)]$\footnote{This is the Iverson
bracket notation, where $[P]$ is $1$ for a predicate $P$ if $P$ is
true and $0$ otherwise.} as an unbiased estimator for
the Jaccard similarity $|A\cap B|/|A\cup B|$.

Unfortunately, we cannot realistically implement perfect minwise hash functions
where each $x\in S$ has probability $1/|S|$ of being the unique minimum
\cite{broder98minwise}. More precisely, to handle any subset $S$ of
a universe $\calU$, we need a random permutation $h:\calU\rightarrow \calU$ represented
using $\Theta(|\calU|)$ bits.

Instead we settle for a bias $\eps$. Formally, a random hash function
$h:\calU\rightarrow \calR$ from some key universe $\calU$ to some range $\calR$ of hash values
is random variable following some distribution over $\calR^\calU$. We
say that $h$ is {\em $\eps$-minwise} or has \emph{bias} $\eps$ if for
every $S\subseteq \calU$ and $x\in \calU\sm S$,
\begin{align}
\Pr[h(x)\leq \min h(S)]\leq \frac{1+\eps}{|S|+1}\label{eq:min-upper}\\
\Pr[h(x)< \min h(S)]\geq \frac{1-\eps}{|S|+1}\label{eq:min-lower}
\end{align}
From \req{eq:min-upper} and \req{eq:min-lower}, we easily get for any
$A,B\subseteq \calU$, that
\[
    \Pr_{h}[\min h(A) = \min h(B)] = (1\pm\eps)\cdot\frac{|A\cap B|}{|A\cup
    B|}\enspace .
\]
To implement $\eps$-minwise hashing in Wegman and Carter's
\cite{wegman81kwise} classic framework of $k$-independent hash
functions $\Theta(\log\frac{1}{\eps})$-independence is both sufficient
\cite{indyk01minwise} and necessary \cite{patrascu10kwise-lb}.  These
results are for ``worst-case'' $k$-independent hash functions. A much more
    time-efficient solution is based on simple tabulation hashing of Zobrist
    \cite{zobrist70hashing}. In simple tabulation hashing, the hash value is
    computed by looking up $c=O(1)$ bitstrings in tables of size
    $|\calU|^{1/c}$ and XORing the results. This is very fast with tables
    in cache. P\v{a}tra\c{s}cu and Thorup have shown \cite{patrascu11charhash}
    that simple tabulation hashing, which is not even $4$-independent, has bias
$\eps=\tilde O(1/|S|^{1/c})$.
Unfortunately, this bias is useless for small sets $S$.

In this paper, we consider the twisted tabulation of P\v{a}tra\c{s}cu and Thorup \cite{PT13:twist} which was invented to yield Chernoff-style concentration bounds, and high probability amortized performance bounds for linear probing. It is almost
as fast as simple tabulation using the same number of lookups but an extra
XOR and a shift. We show that with twisted tabulation, the
bias is $\eps=\tilde O(1/|U|^{1/c})$, which is independent of the set size.

It should be noted, that Thorup \cite{thorup13doubletab} recently
introduced a double tabulation scheme yielding high independence in
$O(1)$ time, hence much faster than using an $\omega(1)$-degree
polynomial to get $\omega(1)$-independence and $o(1)$ bias. However, with table size $|U|^{1/c}$, the
scheme ends up using at least $7c$ lookups \cite[Theorem
  1]{thorup13doubletab} and $12$ times more space, so we expect it to be
at least an order of magnitude slower than twisted tabulation\footnote{The whole area of tabulation hashing
is about minimizing the number of lookups, e.g., \cite{KW12} saves a
factor 2 in lookups over \cite{thorup12kwise} for moderate
independence.}.

When using minwise for similarity estimation, to
reduce variance, we typically want to run $q$ experiments with $q$
independent hash functions $h_1,...,h_q$, and save the vector of $(\min
h_1(A),...,\min h_q(A))$ as a \emph{sketch} for the set $A$. We can then estimate the
Jaccard similarity as $\sum_{i=1}^q [\min h_i(A)=\min h_i(B)]/q$.
While $q$ reduces variance, it does not reduce bias, so the bias has to be
small for each $h_i$. This scheme is commonly referred to as $k\times$minwise.
Since $\min h_1(A)$ is always compared to $\min h_1(B)$, we say
that the samples of the two sketches are \emph{aligned}.
A standard alternative\footnotemark[1], called bottom-$q$, is to just
use a single hash function $h$, and store the $q$ smallest hash
values as a set $S(A)$. Estimating the Jaccard-index is
then done as $|S(A)\cap S(B)\cap \text{\{$q$ smallest values of
$S(A)\cup S(B)$\}}|/q$. It turns out that a large $q$ reduces both variance and bias
\cite{thorup13bottomk}. However, the problem with bottom-$q$ sketches, is that the
samples lose their alignment.
In applications of large-scale machine learning this alignment is needed in order to efficiently construct a dot-product for
use with a linear support vector machine (SVM)
\footnote{See
\url{http://en.wikipedia.org/wiki/Support_vector_machine\#Linear_SVM}}
such as LIBLINEAR
\cite{fan08liblinear} or Pegasos \cite{shwartz07pegasos}. Using the alignment
of $k\times$minwise, it was shown how to construct such a dot-product in
\cite{li11minhash} based on this scheme. In such applications it is
therefore important to have small bias $\eps$.
Finally, we note that when $q=1$, both
schemes reduce to basic minwise hashing with the fundamental goal of sampling
a single random element from any set with only a small bias, which is
exactly the problem addressed in this paper.

\section{Preliminaries}
Let us briefly review tabulation-based hashing. For both simple and twisted
tabulation we are dealing with some universe $\calU = \{0,1,\ldots, u-1\}$
denoted by $[u]$ and
wish to hash keys from $[u]$ into some range $\calR = [2^r]$. We view a key
$x\in [u]$ as a vector of $c>1$ \emph{characters} from the alphabet $\Sigma =
[u^{1/c}]$, i.e.~$x = (x_0,\ldots,x_{c-1})\in \Sigma^c$. We generally assume
$c$ to be a small constant (e.g.~$4$).

\subsection{Simple Tabulation}\label{sec:simple}
In simple tabulation hashing we initialize $c$ tables $h_0, \ldots, h_{c-1} :
\Sigma\to \calR$ with independent random data. The hash $h(x)$ is then computed
as
\[
    h(x) = \bigoplus_{i\in [c]} h_i[x_i]\enspace .
\]
Here $\oplus$ denotes bit-wise XOR. This is a well-known scheme dating back to
\cite{zobrist70hashing}.

Simple Tabulation is known to be $3$-independent, but it was shown in
\cite{patrascu11charhash} to have much more powerful properties than this would
suggest. These properties include fourth moment bounds, Chernoff bounds when
distribution balls into many bins and random graph properties necessary
in cuckoo hashing. It was also shown that simple tabulation is $\eps$-minwise
independent with $\eps = O\left(\frac{\lg^2 n}{n^{1/c}}\right)$.

We will need the following basic lemma regarding simple tabulation (\cite[Lemma
2.2]{patrascu11charhash}):
\begin{lemma}\label{lem:d_bound_bins}
    Suppose we use simple tabulation to hash $n\le m^{1-\eps}$
    keys into $m$ bins for some constant $\eps>0$.
    For any constant $\gamma$, all bins get less than $d =
    \min\{((1+\gamma)/\eps)^c, 2^{(1+\gamma)/\eps}$\} keys with probability
    $\ge 1 - m^{-\gamma}$.
\end{lemma}
Specifically this implies that if we hash $n$ keys into $m = nu^\eps$ bins,
then each bin has $O(1)$ elements with high probability. In this paper ``with
high probability'' (w.h.p.) means with probability $1-u^{-\gamma}$
for any desired constant $\gamma > 1$.

\subsection{Twisted Tabulation}
Twisted tabulation hashing is another tabulation-based hash function
introduced in \cite{PT13:twist}. Twisted tabulation can be seen as two
independent simple tabulation functions $h^\tau : \Sigma^{c-1}\to \Sigma$
and $h^\calS : \Sigma^c\to \calR$. If we view a key $x$ as the head
$head(x) = x_0$ and the tail $tail(x) = (x_1,\ldots, x_{c-1})$, we can
define the hash value of twisted tabulation as follows:
\begin{align*}
    t(x) &= h^\tau(tail(x)) \\
    h_{>0}(x) &= \bigoplus_{i=1}^{c-1} h_i^\calS[x_i] \\
    h(x) &= h_{>0}(x)\xor h^\calS_0[x_0\xor t(x)]\enspace .
\end{align*}
We refer to the value $x_0\xor t(x)$ as the \emph{twisted head} of the key
$x$, and define the \emph{twisted group} of a character $\alpha$ to be
$G_\alpha = \{x\cond x_0\xor t(x) = \alpha\}$. For the keys in $G_\alpha$, we
refer to the XOR with $h^\calS_0[x_0\xor t(x)]$ as the \emph{final
(XOR)-shift}, which is common to all keys in $G_\alpha$. We call $h_{>0}(x)$ the
\emph{internal hashing}.

Throughout the proofs we will rely on the independence between $h^\tau$
and $h^\calS$ to fix the hash function in a specific order, i.e.~fixing the
twisted groups first.

One powerful property of twisted tabulation is that the keys are
distributed nicely into the twisted groups. We will use the following lemma from
the analysis of twisted tabulation \cite[Lemma 2.1]{PT13:twist}:

\begin{lemma}\label{lem:twistgroups}
    Consider an arbitrary set $S$ of keys and a constant parameter $\eps > 0$.
    W.h.p.~over the random choice of the twister hash function, $h^\tau$, all twisted
    groups have size $O(1 + |S|/\SigEps)$.
\end{lemma}

Twisted tabulation hashing also gives good concentration bounds in form of
Chernoff-like tail bounds, which is captured by the following lemma,
\cite[Theorem 1.1]{PT13:twist}.

\begin{lemma}\label{lem:tailbounds}
    Choose a random twisted tabulation hash function $h : [u]\to [u]$. For each
    key $x\in [u]$ in the universe, we have an arbitrary \emph{value function}
    $v_x : [u]\to [0,1]$ assigning a value $V_x = v_x(h(x))\in [0,1]$ to $x$
    for each possible hash value. Let $\mu_x = \E_{y\in [u]}[v_x(y)]$ denote the
    expected value of $v_x(y)$ for uniformly distributed $y\in [u]$. For a fixed
    set of keys $S\subseteq [u]$, define $V = \sum_{x\in S} V_x$ and
    $\mu = \sum_{x\in S} \mu_x$. Let $\gamma$, $c$, and $\eps$ be constants.
    Then for any $\mu < \SigEps$ and $\delta > 0$ we have:
    \begin{align}
        \Prp{V\ge (1+\delta)\mu} &\le
        \left(\frac{e^\delta}{(1+\delta)^{(1+\delta)}}\right)^{\Omega(\mu)} +
        1/u^\gamma \\
        \Prp{V\le (1-\delta)\mu} &\le
        \left(\frac{e^{-\delta}}{(1-\delta)^{(1-\delta)}}\right)^{\Omega(\mu)} +
        1/u^\gamma
    \end{align}
\end{lemma}

In practice, we can merge $h^\tau$ and $h^\calS$ to a single simple
tabulation function $h^\star : \Sigma\to \Sigma\times \calR$, but with
$h^\star_0 : \Sigma\to \calR$. This adds $\log \Sigma$ bits to each entry
of the tables $h^\star_1,\ldots h_{c-1}^\star$ (in practice we want these
to be 32 or 64 bits anyway). See the code in \autoref{fig:twisted_code} for an
implementation of 32-bit keys in C.

\begin{figure}[htbp]
\begin{verbatim}
INT32 TwistedTab32(INT32 x, INT64[4][256] H) {
    INT32 i;
    INT64 h=0;
    INT8 c;
    for (i=0;i<3;i++) {
      c=x;
      h^=H[i][c];
      x = x>> 8;
    }                         // at the end i=3
    c=x^h;                    // extra xor with h
    h^=H[i][c];
    h>>=32;                   // extra shift of h
    return ((INT32) h);
}
\end{verbatim}
\caption{C-code implementation of twisted tabulation for 32-bit keys assuming a
point H to randomly fille storage.}
\label{fig:twisted_code}
\end{figure}

\section{Minwise for twisted tabulation}
We will now show the following theorem:
\begin{theorem}\label{thm:minwise_twist}
    Twisted tabulation is $O\!\left(\frac{\log^2 u}{\Sigma}\right)$-minwise independent.
\end{theorem}

Recall from the definition of $\eps$-minwise, that we are given an input
set $S$ of $|S| = n$ keys and a query key $q\in \calU\sm S$. We will denote by
$Q$ the twisted group of the query key $q$.
Similarly to the analysis in \cite{patrascu11charhash} we assume that the output
range is $[0,1)$. We pick $\ell = \gamma\log u$ and divide the
output range into $n/\ell$ bins. Here $\gamma$ is chosen such that the number
of bins is a power of two and large enough that the following two properties hold.
\begin{enumerate}
    \item The minimum bin $[0,\ell/n)$ is non-empty with probability $1-1/u^2$
            by \autoref{lem:tailbounds}. Here $\mu = O(\log u) < \SigEps$.
    \item The bins are $d$-bounded for each twisted group (for some constant
        $d$) with probability $1-1/u^2$ by \autoref{lem:d_bound_bins}.
        Meaning that for any twisted group $G$, at most $d$ keys land in
        each of the $n/\ell$ bins after the internal hashing is done.
        This holds because each twisted group has $n/\SigEps$ elements
        w.h.p.~by \autoref{lem:twistgroups}.
\end{enumerate}

Similar to \cite{patrascu11charhash}, we assume that the hash values
are binary fractions of infinite precision so we can ignore collisions. The
theorem holds even if we use just $\lg(n\Sigma)$ bits for the
representation: Let $\tilde{h}$ be the truncation of $h$ to $\lg(n\Sigma)$
bits. There is only a distinction when $\tilde{h}(q)$ is
minimal and there exists some $x\in S$ such that $\tilde{h}(x) = \tilde{h}(q)$.
Since the minimum bin is non-empty with probability $1-1/u^2$ we can bound the
probability of this from above by
\[
    \Prp{\tilde{h}(q)\le \ell/n\land \exists x\in S : \tilde{h}(x) =
    \tilde{h}(q)}
    \le \frac{\ell}{n}\cdot\left(n\cdot \frac{1}{n\Sigma}\right) + 1/u^2
\]
using 2-independence to conclude that $\{\tilde{h}(q)\le \ell/n\}$ and
$\{\tilde{h}(x) = \tilde{h}(q)\}$ are independent.

\subsection{Upper bound}
To upper bound the probability that $h(q)$ is smaller than $\min
h(S)$ it suffices to look at the case when $q$ is in
the minimum bin $[0,\ell/n)$, as we have
\begin{align*}
    \Prp{h(q) < \min h(S)}
    &\le \Prp{\min h(S)\ge\ell/n} + \Prp{h(q)<\min(h(S)\cup \{\ell/n\})} \\
    &\le 1/u^2 + \Prp{h(q) < \min(h(S)\cup \{\ell/n\})} \numberthis
    \label{eq:ub_main}
\end{align*}

To bound \eqref{eq:ub_main} we will use the same notion of
representatives as in \cite{patrascu11charhash}: If a non-query twisted group
$G_\alpha\ne Q$ has more than one element in some bin, we pick one of these
arbitrarily as
the representative. Let $R(G_\alpha)$ denote the set of representatives from
$G_\alpha$ and let $R$ denote the union of all such sets. We trivially
have that $\Prp{h(q) < \min h(S)} \le \Prp{h(q) < \min h(R)}$.

The proof relies on fixing the tables associated with the hash functions
$h^\tau$ and $h^\mathcal{S}$ in the following order:
\begin{enumerate}
    \item Grouping into twisted groups is done by fixing $h^\tau$. Each group
        has $O(1+n/\SigEps)$ elements by \autoref{lem:twistgroups} w.h.p.
    \item The internal hashing of all twisted groups is done by fixing the
        tables $h_1^\mathcal{S},\ldots,h_{c-1}^\mathcal{S}$. This
        determines the set of representatives $R$.
    \item Having fixed the set $R$ we do the final shifts of the twisted groups
        $G_\alpha$ by fixing $h_0^\mathcal{S}$. We will show that the
        probability of $q$ having the minimum
        hash value after these shifts is at most $1/(|R|+1)$.

        Since $|R|$ is a random variable depending only on the internal hashing
        and twisted groups, the entire probability is bounded by $\Ep{1/(|R|+1)}$.
\end{enumerate}

To see step 3 from above we let $Rand(A)$ be a randomizing function that
takes each element in a set $A$ and replaces it with an independent uniformly
random number in $[0,1)$. We will argue that
\begin{equation}\label{eq:r_rand}
    \Prp{h(q) < \min h(R)\cup\{\ell/n\}} \le \Prp{h(q) < \min Rand(R)} =
    1/(|R|+1)
\end{equation}

To prove \eqref{eq:r_rand} fix $h(q) = p<\ell/n$ and consider some twisted
group $G_\alpha$. When doing the final shift of the group we note that each
representative $x\in R(G_\alpha)$ is shifted randomly, so $\Prp{h(x) \le p} = p$.
However, since the number of bins is a power of two, and each
representative in $R(G_\alpha)$ is shifted by the same value, at most one
element of $R(G_\alpha)$ can land in the minimum bin. This gives
$\Prp{\min h(R(G_\alpha))\le p} = |R(G_\alpha)|p$. For $Rand(R)$, a union bound
gives that $\Prp{\min Rand(R(G_\alpha))\le p}\le |R(G_\alpha)|p$, implying that
\[
    \Prp{p<\min (h(R(G_\alpha))\cup\{\ell/n\})} \le
    \Prp{p<\min (Rand(R(G_\alpha))\cup\{\ell/n\})}
\]
Because the shifts of different twisted groups are done independently we get
\begin{align*}
    \Prp{p < \min(h(R)\cup\{\ell/n\})}
    &= \prod_{G_\alpha\ne Q} \Prp{p < \min( h(R(G_\alpha))\cup\{\ell/n\})} \\
    &\le \prod_{G_\alpha\ne Q} \Prp{p < \min (Rand(R(G_\alpha))\cup\{\ell/n\})} \\
    &= \Prp{p < \min (Rand(R)\cup\{\ell/n\})} \\
    &\le \Prp{p < \min Rand(R)} \\
\end{align*}
This holds for any value $p< \ell/n$, so it also holds for our random hash
value $h(q)$. Therefore
\[
    \Prp{h(q) < \min(h(R)\cup\{\ell/n\})} \le \Prp{h(q) < \min Rand(R)}\le 1/(|R|+1)
\]
This finishes the proof of \eqref{eq:r_rand}.

All that remains is to bound the expected value $\Ep{1/(|R|+1)}$ and thus the
total probability when the internal hashing and twisted groups are random.
We will do this using a convexity argument, so we need the following
constraints on the random variable $|R|$:
We trivially have $1\le |R|\le n$. We know that the internal hashing is
$d$-bounded with probability $1-1/u^2$, which gives $|R|\ge |S\sm Q|/d\ge
n/(2d)$. To bound $\Ep{|R|}$ from below, consider the probability that a key
$x$ is \emph{not} a representative. For this to happen $x$ must land in the
query group, or another element must land in the same
twisted group and bin as $x$. By 2-independence and a union bound the probability
of this event is at most $1/\Sigma + (n-1)\cdot 1/\Sigma\cdot \ell/n =O(\ell/\Sigma)$.
The expected number of representatives is therefore
\begin{align*}
    \Ep{|R|} &= \sum_{x\in S}\Prp{x\in R} \\
             &= \sum_{x\in S}(1 - \Prp{x\notin R}) \\
             &\ge n\cdot (1 - O(\ell/\Sigma))\enspace .
\end{align*}

To bound $\Ep{1/(|R| + 1)}$ we introduce a random variable $r$ which maximizes
$\Ep{1/(r+1)}$ while satisfying the constraints of $|R|$ noted above. By convexity
of $1/(r+1)$ we get that $\Ep{1/(r+1)}$ is maximized when $r$ takes the most extreme
values. Hence $r=1$ with probability $1/u^2$, $r = n/(2d)$ with the
maximal probability $p$ and $r = n$ with probability $(1-p-1/u^2)$.
This gives an expected value of
\[
    \Ep{r} = 1/u^2 + p\cdot n/(2d) + (1 - p - 1/u^2)\cdot n\enspace .
\]
Thus $p = O(\ell/\Sigma)$ to respect the constraints. To bound
$\Ep{1/(|R|+1)}$ we have
\begin{align*}
    \Ep{1/(|R|+1)}
    &\le \Ep{1/(r+1)} \\
    &\le \frac{1}{2u^2} + \frac{p}{n/(2d) +1} + \frac{1-p-1/u^2}{n+1} \\
    &\le \frac{O(p)}{n+1} + \frac{1}{n+1} + O(1/u^2) \\
    &= \frac{1}{n+1}\cdot (1 + O(\ell/\Sigma))\enspace . \numberthis\label{eq:exp_rinv}
\end{align*}

Combining \eqref{eq:ub_main}, \eqref{eq:r_rand} and \eqref{eq:exp_rinv} we get
\begin{align*}
    \Prp{h(q) < \min h(S)}
    &\le \Prp{h(q) < \min(h(S)\cup\{\ell/n\})} + O(1/u^2) \\
    &\le \Prp{h(q) < \min(h(R)\cup\{\ell/n\})} + O(1/u^2) \\
    &\le \Ep{1/(|R|+1)} + O(1/u^2) \\
    &= \frac{1}{n+1}\cdot\left(1 + O\!\left(\frac{\log
    u}{\Sigma}\right)\right)\enspace .
\end{align*}

\subsection{Lower bound}
We have two cases for the lower bound. When $n = O(\log u)$ we observe
that the probability of some twisted group having more than one element is
bounded from above by $n^2/\Sigma = O(\log^2 u/\Sigma)$ using 2-independence
and a union bound. Since the twisted groups hash independently of each other
we have in this case that all elements hash independently. The probability of
$q$ getting the smallest hash value is thus at least
$1/(n+1)\cdot(1-O(\log^2 u/\Sigma))$.

When $n=\omega(\log u)$ we again look at the case when $q$
lands in the minimum bin $[0,\ell/n)$. We consider the query group $Q$
separately and thus look at the expression:
\begin{align*}
    \Prp{h(q) < \min h(S)} &\ge \Prp{h(q) < \min(h(S) \cup\{\ell/n\})} \\
                           &= \Prp{h(q) < \min(h(S\sm Q)\cup \{\ell/n\})} \\
                           &\quad-\Prp{\min h(Q) < h(q) < \min(h(S\sm
Q)\cup\{\ell/n\})}\enspace . \numberthis \label{eq:lb_noquery}
\end{align*}
Furthermore we will assume that all twisted groups have
$O(1+n/\Sigma^{1-\eps})$ elements at the cost of a factor $(1 - 1/u^2)$ by
\autoref{lem:twistgroups}. We will subtract this extra term later in
\eqref{eq:lb_part1done}. Since the twisted groups hash independently we have
for a fixed $h(q) = p < \ell/n$ that
\begin{equation}\label{eq:lb_prod}
    \Prp{p < \min h(S\sm Q)} = \prod_{G_\alpha\ne Q}\Prp{p < \min
    h(G_\alpha)}\enspace .
\end{equation}
We can bound this expression using \cite[Lemma 5.1]{patrascu11charhash}, which
states that $1 - pk > (1-p)^{(1+pk)k}$ for $pk \le \sqrt{2}-1$ and $p\in[0,1]$.
Consider a twisted group $G_\alpha$ and some element $x\in G_\alpha$. We have
$\Prp{h(x) < p} = p$ and a union bound gives us that $\Prp{p < \min
h(G_\alpha)}\ge 1 - p|G_\alpha|$. Since $n = \omega(\log u)$ we have
that $p|G_\alpha| \le \ell/n\cdot O(1 + n/\Sigma^{1-\eps}) = o(1)$, so the
conditions for the lemma hold. This gives us
\begin{equation}\label{eq:lb_group}
    1 - p|G_\alpha| \ge (1-p)^{|G_\alpha|(1+p\cdot |G_\alpha|)}
\end{equation}
Plugging this into \eqref{eq:lb_prod} gives
\begin{align*}
    \Prp{p < \min h(S\sm Q)}
    &\ge \prod_{G_\alpha\ne Q} (1-p)^{|G_\alpha|(1+p|G_\alpha|)} \\
    &\ge \prod_{G_\alpha\ne Q} (1-p)^{|G_\alpha|(1+p\cdot 2(|G_\alpha|-1))} \\
    &\ge (1-p)^m,
\end{align*}
with
\[
    m = n + O(\ell/n)\cdot\sum_{G_\alpha\ne Q} (|G_\alpha| - 1)|G_\alpha|\enspace .
\]
To bound the entire probability we thus integrate from $0$ to $\ell/n$:
\begin{align*}
    \Prp{h(q)\le \min(h(S\sm Q)\cup\{\ell/n\})}
    &= \int_0^{\ell/n} \Prp{p < \min h(S\sm Q)} \mathrm{d} p \\
    &\ge \int_{0}^{\ell/n} (1-p)^m \mathrm{d} p \\
    &\ge \frac{1 - (1-\ell/n)^{m+1}}{m+1} \\
    &>1/(m+1) - 1/(nu)\enspace .\numberthis\label{eq:lb_part1}
\end{align*}
Similar to the upper bound $m$ only depends on the twisted groups and their
internal hashing, so the entire probability is bounded by
$\Ep{1/(m+1)}-1/nu\ge 1/\Ep{m+1} - 1/nu$. We note that the sum $\sum_{G_\alpha\ne Q}
(|G_\alpha| - 1)|G_\alpha|$ counts for each key in a non-query group the number
of other elements in its group, so
\[
    \Ep{\sum_{G_\alpha\ne Q} (|G_\alpha|-1)|G_\alpha|} \le n^2/\Sigma\enspace .
\]
The expected value $\Ep{m+1}$ is therefore bounded by
\begin{equation}\label{eq:lb_expm}
    \Ep{m+1} \le (n+1)\cdot (1 + O(\ell/\Sigma))\enspace .
\end{equation}
We can combine this with \eqref{eq:lb_part1} and get a bound on the first part
of \eqref{eq:lb_noquery}. We also need to subtract the probability that the
keys don't distribute nicely into twisted groups. Doing this we get the
following bound:
\begin{align*}
    \Prp{h(q)\le \min(h(S\sm Q)\cup \{\ell/n\})}
    &\ge \Ep{1/(m+1)} - 1/nu - 1/u^2 \\
    &\ge 1/\Ep{m+1} - 1/nu -1/u^2 \\
    &\ge \frac{1}{(n+1)(1 + O(\ell/\Sigma))} - 1/nu - 1/u^2 \\
    &\ge \frac{1}{n+1} \cdot \left(1 -
        O\left(\frac{\log u}{\Sigma}\right)\right)
    \numberthis\label{eq:lb_part1done}
\end{align*}

To finish the bound on \eqref{eq:lb_noquery} we need to give an upper bound on
\begin{equation}\label{eq:lb_part2}
    \Prp{\min h(Q) < h(q) < \min h(S\sm Q)\land h(q) < \ell/n}\enspace .
\end{equation}
To do this we will again consider the set of representatives that we used in
the upper bound. We start by fixing the twisted groups. Just like in the upper
bound we have w.h.p.~that $|R|\ge n/(2d)$. We can therefore bound
\eqref{eq:lb_part2} by
\[
    1/u^2 + \Prp{\min h(Q) < h(q) < \min h(S\sm Q)\land h(q)< \ell/n
    \land |R| \ge n/(2d)}\enspace .
\]
We fix $h(q) = p$ for some $p < \ell/n$. Using 2-independence between the fixed
query value $p$ and each element of $Q$ we get $\Prp{\min h(Q)< p}\le p|Q|$ and
thus
\begin{equation}\label{eq:min_q}
    \Prp{\min h(Q) < p \land |R|\ge n/(2d)} \le p|Q|\enspace .
\end{equation}
We wish to multiply this by
\[
    \Prp{p < \min h(S\sm Q)\ |\ \min h(Q) < p \land |R|\ge n/(2d)}\enspace .
\]
For this we use the same approach as for \eqref{eq:r_rand}. We know that when
$p < \ell/n$ we have that
$\Prp{p < \min h(R)} \le \Prp{p < \min Rand(R)} = (1-p)^{|R|}$. This holds
regardless of the internal hashing so our restriction of $|R|\ge n/(2d)$ does
not change anything. We now get
\[
    \Prp{p < \min h(S\sm Q)\ |\ \min h(Q) < p \land |R|\ge n/(2d)} \le
    (1-p)^{n/(2d)}\enspace .
\]
Multiplying together with \eqref{eq:min_q} we get
\begin{align*}
    \Prp{\min h(Q) < p < \min h(S\sm Q)\land |R|\ge n/(2d)}
    &\le p|Q|(1-p)^{n/(2d)}\\
    &\le p|Q|e^{-pn/(2d)}
\end{align*}
for a fixed $p<\ell/n$. To finish the bound we thus integrate from $0$ to
$\ell/n$ and get an upper bound on \eqref{eq:lb_part2}:
\begin{align*}
    &\Prp{\min h(Q) < h(q) < \min h(S\sm Q)\land h(q) < \ell/n}\\
    &\qquad\le 1/u^2 + \Prp{\min h(Q) < h(q) < \min h(S\sm Q)\land
        h(q) < \ell/n \land |R|\ge n/(2d)} \\
    &\qquad\le1/u^2 + \int_0^{\ell/n} p|Q|e^{-pn/(2d)}\ dp \\
    &\qquad =1/u^2 + O\left(\int_0^{d/n}p|Q|\ dp\right) \\
    &\qquad =1/u^2 + O(|Q|/n^2)\enspace .
\end{align*}
We now note that $|Q|$ is a random variable with expected value $n/\Sigma$,
which gives the final bound on \eqref{eq:lb_part2} as
\begin{align*}
    \Prp{\min h(Q) < h(q) < \min h(S\sm Q)\land h(q) < \ell/n}
    &\le \Ep{O(|Q|/n^2)} \\
    &= O(1/n\Sigma)\enspace .\numberthis\label{eq:lb_querybound}
\end{align*}

Combining \eqref{eq:lb_noquery}, \eqref{eq:lb_part1done} and
\eqref{eq:lb_querybound} gives the desired bound:
\begin{align*}
    \Prp{h(q) < \min h(S)}
    &\ge \Prp{h(q) < \min h(S\sm Q) \land h(q) < \ell/n} \\
    &\quad-\Prp{\min h(Q) < h(q) < \min h(S\sm Q) \land h(q) < \ell/n}\\
    &\ge \frac{1}{n+1}\cdot\left(1 -
        O\left(\frac{\log u}{\Sigma}\right)\right) -
        O\left(\frac{1}{n\Sigma}\right) \\
    &= \frac{1}{n+1}\cdot\left(1 -
        O\left(\frac{\log u}{\Sigma}\right)\right)
\end{align*}

\bibliographystyle{amsplain}
\bibliography{general}

\providecommand{\bysame}{\leavevmode\hbox to3em{\hrulefill}\thinspace}
\providecommand{\MR}{\relax\ifhmode\unskip\space\fi MR }
\providecommand{\MRhref}[2]{%
  \href{http://www.ams.org/mathscinet-getitem?mr=#1}{#2}
}
\providecommand{\href}[2]{#2}
\begin{thebibliography}{10}

\bibitem{BHP09}
Yoram Bachrach, Ralf Herbrich, and Ely Porat, \emph{Sketching algorithms for
  approximating rank correlations in collaborative filtering systems}, Proc.
  16th SPIRE, 2009, pp.~344--352.

\bibitem{BPR09}
Yoram Bachrach, Ely Porat, and Jeffrey~S. Rosenschein, \emph{Sketching
  techniques for collaborative filtering}, Proc. 21st IJCAI, 2009,
  pp.~2016--2021.

\bibitem{broder97onthe}
Andrei~Z. Broder, \emph{On the resemblance and containment of documents}, Proc.
  Compression and Complexity of Sequences (SEQUENCES), 1997, pp.~21--29.

\bibitem{Broder00}
Andrei~Z. Broder, \emph{Identifying and filtering near-duplicate documents},
  Proc. 11th CPM, 2000, pp.~1--10.

\bibitem{broder98minwise}
Andrei~Z. Broder, Moses Charikar, Alan~M. Frieze, and Michael Mitzenmacher,
  \emph{Min-wise independent permutations}, Journal of Computer and System
  Sciences \textbf{60} (2000), no.~3, 630--659, See also STOC'98.

\bibitem{broder97minwise}
Andrei~Z. Broder, Steven~C. Glassman, Mark~S. Manasse, and Geoffrey Zweig,
  \emph{Syntactic clustering of the web}, Computer Networks \textbf{29} (1997),
  1157--1166.

\bibitem{CDFGIMUY01}
Edith Cohen, Mayur Datar, Shinji Fujiwara, Aristides Gionis, Piotr Indyk,
  Rajeev Motwani, Jeffrey~D. Ullman, and Cheng Yang, \emph{Finding interesting
  associations without support pruning}, IEEE Trans. Knowl. Data Eng.
  \textbf{13} (2001), no.~1, 64--78.

\bibitem{DatarM02estimatingrarity}
Mayur Datar and S.~Muthukrishnan, \emph{Estimating rarity and similarity over
  data stream windows}, Proc. 10th ESA, 2002, pp.~323--334.

\bibitem{fan08liblinear}
Rong-En Fan, Kai-Wei Chang, Cho-Jui Hsieh, Xiang-Rui Wang, and Chih-Jen Lin,
  \emph{{LIBLINEAR}: A library for large linear classification}, Journal of
  Machine Learning Research \textbf{9} (2008), 1871--1874.

\bibitem{henzinger06duplicates}
Monika~Rauch Henzinger, \emph{Finding near-duplicate web pages: a large-scale
  evaluation of algorithms}, Proc. ACM SIGIR, 2006, pp.~284--291.

\bibitem{indyk01minwise}
Piotr Indyk, \emph{A small approximately min-wise independent family of hash
  functions}, Journal of Algorithms \textbf{38} (2001), no.~1, 84--90, See also
  SODA'99.

\bibitem{KW12}
Toryn~Qwyllyn Klassen and Philipp Woelfel, \emph{Independence of
  tabulation-based hash classes}, Proc. 10th Latin American Theoretical
  Informatics (LATIN), 2012, pp.~506--517.

\bibitem{li11minhash}
Ping Li, Anshumali Shrivastava, Joshua~L. Moore, and Arnd~Christian K{\"o}nig,
  \emph{Hashing algorithms for large-scale learning}, Advances in Neural
  Information Processing Systems, 2011, pp.~2672--2680.

\bibitem{manku07duplicates}
Gurmeet~Singh Manku, Arvind Jain, and Anish~Das Sarma, \emph{Detecting
  near-duplicates for web crawling}, Proc. 10th WWW, 2007, pp.~141--150.

\bibitem{patrascu10kwise-lb}
Mihai P{\v a}tra{\c s}cu and Mikkel Thorup, \emph{On the $k$-independence
  required by linear probing and minwise independence}, Proc. 37th
  International Colloquium on Automata, Languages and Programming (ICALP),
  2010, pp.~715--726.

\bibitem{patrascu11charhash}
\bysame, \emph{The power of simple tabulation-based hashing}, Journal of the
  ACM \textbf{59} (2012), no.~3, Article 14, Announced at STOC'11.

\bibitem{PT13:twist}
Mihai P\v{a}tra\c{s}cu and Mikkel Thorup, \emph{Twisted tabulation hashing},
  Proc. 24th ACM/SIAM Symposium on Discrete Algorithms (SODA), 2013,
  pp.~209--228.

\bibitem{SWA03}
Saul Schleimer, Daniel~Shawcross Wilkerson, and Alexander Aiken,
  \emph{Winnowing: Local algorithms for document fingerprinting}, Proc. SIGMOD,
  2003, pp.~76--85.

\bibitem{shwartz07pegasos}
Shai Shalev-Shwartz, Yoram Singer, and Nathan Srebro, \emph{Pegasos: Primal
  estimated sub-gradient solver for svm}, Proceedings of the 24th International
  Conference on Machine Learning, ICML '07, 2007, pp.~807--814.

\bibitem{thorup13bottomk}
Mikkel Thorup, \emph{Bottom-k and priority sampling, set similarity and subset
  sums with minimal independence}, Proc. 45th ACM Symposium on Theory of
  Computing (STOC), 2013.

\bibitem{thorup13doubletab}
\bysame, \emph{Simple tabulation, fast expanders, double tabulation, and high
  independence}, FOCS, 2013, pp.~90--99.

\bibitem{thorup12kwise}
Mikkel Thorup and Yin Zhang, \emph{Tabulation-based 5-independent hashing with
  applications to linear probing and second moment estimation}, SIAM Journal on
  Computing \textbf{41} (2012), no.~2, 293--331, Announced at SODA'04 and
  ALENEX'10.

\bibitem{wegman81kwise}
Mark~N. Wegman and Larry Carter, \emph{New classes and applications of hash
  functions}, Journal of Computer and System Sciences \textbf{22} (1981),
  no.~3, 265--279, See also FOCS'79.

\bibitem{YC06}
Hui Yang and James~P. Callan, \emph{Near-duplicate detection by instance-level
  constrained clustering}, Proc. 29th SIGIR, 2006, pp.~421--428.

\bibitem{zobrist70hashing}
Albert~Lindsey Zobrist, \emph{A new hashing method with application for game
  playing}, Tech. Report~88, Computer Sciences Department, University of
  Wisconsin, Madison, Wisconsin, 1970.

\end{thebibliography}

\end{document}